\documentclass[aps,prl,twocolumn,showpacs,floatfix,amssymb,amsmath]{revtex4}
\usepackage{graphicx}
\usepackage{dcolumn}
\usepackage{bm}
\usepackage{psfrag}
\usepackage{color}
\newcommand{\be}{\begin{equation}}
\newcommand{\ee}{\end{equation}}
\newcommand{\bea}{\begin{eqnarray}}
\newcommand{\eea}{\end{eqnarray}}
\newcommand{\p}{\partial}
\newcommand{\s}{\sigma}

\newcommand{\re}{\mbox{e}}
\newcommand{\ba}{\begin{array}}
\newcommand{\ea}{\end{array}}
\def\nn{\nonumber\\}

\begin{document}

\title{Electron-electron and spin-orbit interactions in armchair graphene ribbons}

\author{Mahdi Zarea} 
\author{Nancy Sandler} 
\affiliation{Dept. of Physics and Astronomy, Nanoscale
and Quantum Phenomena Institute \\Ohio University, Athens, Ohio
45701-2979}

\date{\today}

\begin{abstract}

The effects of intrinsic spin-orbit and Coulomb interactions on low-energy properties of finite width graphene armchair ribbons are studied by means of a Dirac Hamiltonian. It is shown that metallic states subsist in the presence of intrinsic spin-orbit interactions as spin-filtered edge states, in contrast with the insulating behavior predicted for graphene planes. A charge-gap opens due to Coulomb interactions in neutral ribbons, that vanishes as $\Delta \sim 1/W$, with a gapless spin sector. Weak intrinsic spin-orbit interactions do not change the insulating behavior. Explicit expressions for the width-dependent gap and various correlation functions are presented.  
\end{abstract}

\pacs{73.22.-f, 72.80.Rj, 75.70.Ak, 75.10.Pq} 
\maketitle
Since the discovery of the anomalous quantum Hall effect in graphene \cite{Gaim, Kim}, the single-layer carbon material has received ever-growing attention due to its unusual physical properties and potential technological applications. The unsual properties stem from the peculiar band-structure of the material, that at low-energies, can be described by a Dirac-type Hamiltonian for massless electrons  \cite{physicstoday}. For this reason, much of the phenomena associated with QED can be studied for the first time in a condensed matter system in a controlled manner, even at room temperatures. At the same time, as graphene holds the promise to revolutionize the future of electronics, increasing efforts are made to obtain samples with tailored properties. In this regard, it is remarkable that in the last year, two groups have already succeeded in fabricating the first set of graphene ribbons with variable width \cite{kim2, avouris} for systematic studies. As a consequence, it is essential to understand the properties of the material in confined geometries as controlled production of graphene nanostructures is becoming a reality.
Moreover, it would also be interesting to understand the confinement of Dirac fermions in the presence of interactions as it may have relevant consequences in other areas of physics.

The purpose of this paper is to provide precise answers to two key questions that arise when confinement effects are important: a) what are the consequences of the enhanced electron-electron interaction on transport properties of graphene ribbons? and b) are there metallic graphene nanoribbons in the presence of intrinsic spin-orbit (I-SO) interactions? These two issues have received much attention in the cases of two-dimensional graphene planes \cite{kanemele,yao} and rolled graphene (nanotubes) \cite{fisher, gogolin}, but no detailed study exists at present for nanoribbons with defined edges. 
In this work, we predict that small-momentum scattering introduced by electron-electron interactions opens a width-dependent charge gap in the spectrum of half-filled ribbons, that vanishes in the limit of wide ribbons. We also present evidence for an incipient magnetic order in narrow ribbons, in agreement with recent experimental measurements \cite{abanin}, and show that these results are minimally affected by a weak I-SO interaction. Unlike previous numerical results \cite{son}, the approach presented here gives insight into the processes that originate the gap, allowing a systematic account of the various scattering processes due to interactions, and provides expressions for several correlation functions. Furthermore, we demonstrate that metallic states exist in armchair ribbons of special widths {\it even} in the presence of I-SO interactions and provide expressions for the associated wave functions. This is a fundamental difference from reported results on graphene sheets where insulating behavior was predicted \cite{yao}.

A generic graphene ribbon has a combination of two types of edge terminations: armchair and zigzag. Both edges are associated with characteristic transport behavior in tight-binding calculations: zigzag ribbons are predicted to be metallic due to a topological edge state; and armchair ribbons are predicted to be metallic or insulating depending on the ribbon's width \cite{fujita, nakada, kanemele, brey}. In what follows we focus on finite width armchair ribbons.

We describe a graphene ribbon using an hexagonal Bravais lattice with a unit cell containing two carbon atoms $A$ and $B$. An atom $A$ is connected to its nearest neighbor by $\vec{\delta}_{1} = a(0,1/\sqrt{3}), \vec{\delta}_{2} = (a/2)(-1,-1/\sqrt{3}), \vec{\delta}_{3} = (a/2)(1,-1/\sqrt{3})$; with the lattice unit vectors given by: $\vec{a_{i}} = (1/2) \epsilon_{ijk}(\vec{\delta_{j}} - \vec{\delta_{k}})$. 
The simplest model for electrons in graphene is given in terms of a nearest-neighbor hopping tight-binding Hamiltonian $ 
H_0=\sum_{<ij>}t{c}_{i}^{\dag}{c}_j+h.c$; where $i,j$ label position and spin degrees of freedom in real space. The resulting band-structure contains six equivalent points in the Brillouin zone with vanishing density of states, from which only two are independent (the rest are obtained from reciprocal lattice vector translations). We choose these two independent points to be located at: $K, K'=\pm{4\pi\over3a}(1,0)$. We work with a spin-dependent basis, the \emph{pseudo-spin basis} $\Psi^{\dagger, s}=\left( u^s_A; u^s_B \right)$, where the first (second) component represents the amplitude of the wavefunction at a lattice $A (B)$ site, and $s$ labels the spin. In momentum space the Hamiltonian takes the form:
\begin{eqnarray}
&&H_0=t\left( \begin{array}{cc}
                              0                  & \phi\\
			      \phi^{\star} & 0  \end{array}\right),
\end{eqnarray}
where $\phi=1+2\cos(q_xa/2)e^{-i\sqrt{3}k_ya/2}$ with a spectrum $E_{\pm}(\vec{k})= \pm t|\phi(\vec{k})|$.	
At low energies, the Hamiltonian can be reduced even further to its Dirac form, by expanding $\phi(\vec{k})$ around the $K, K'$ points:
\begin{eqnarray}
&&H_0=v\left( \begin{array}{cc}
                              0                  & ik_y\mp k_x\\
			      -ik_y\mp k_x & 0  \end{array}\right)
\end{eqnarray}
with $v=t\sqrt{3}/2$ (we use $\hbar=1$) and $k_x=q_x-K$.    
Consider a ribbon of length $L$ along the $y$ direction with finite width $W$ along $x$. As shown in \cite{brey, tw}, the wave function that vanishes at the boundaries $x=0, W'=W+a/2$, contains two states at $(K \pm k_{x})$ and $(K' \pm k_{x})$
where $k_x$ takes the values $(k_x-{\pi\over3})W'=\pi n$. The corresponding energy is given by $E=\pm v\sqrt{k_y^2+k_x^2}$. For ribbon widths $W=(3M+1)a$, a set of linear dispersion ($E=\pm vk_y$) states appears with wave functions given by:
\bea
&&\Psi_{\mp}^s=\sqrt{\frac{2}{W'}}\sin\big(\frac{4\pi x}{3a}\big){\frac{\re^{ik_yy}}{\sqrt{2L}}}\left( \begin{array}{c}
                              1\\
			      \mp i    \end{array}\right) \; . \label{freewavefunction}
\eea
This solution defines left ($\Psi_{r=+}$) and right movers ($\Psi_{r=-}$)
for each spin component. At energies below the bulk band-gap (of the order of $v_F/W\approx 0.4 eV$ for $W/a = 10$), these linear modes dominate the physics and they are the only ones considered in the rest of the paper.

{\it Effect of Coulomb interactions.} 
It is known that Coulomb interactions produce drastic changes in the ground state of carbon nanotubes resulting in Luttinger liquid physics at low-energies \cite{yoshi1, fisher}. We show next that they have similar effects for narrow armchair ribbons.

The unscreened Coulomb interaction in two-dimensional graphene is given by $U(x,y)=e^2/(\kappa\sqrt{a_0^2+x^2+y^2})$
where $a_0\approx a/2$ is the radius of carbon $p_z$ orbitals  and we take the value of
$\kappa \simeq 2.45$ appropriate for a graphene sheet on top of an insulating substrate \cite{gogolin}. The interacting Hamiltonian is $H_{int}=(1/2)\int dy dy' { \cal H}$ where: 
\bea
{\cal H}= \sum_{pp',ss'}V_{pp'}(y-y')\psi^{\dag,s}_{p}(y)\psi^{\dag,s'}_{p'}(y')\psi^{s'}_{p'}(y')\psi^{s}_{p}(y)\,&&
\eea
with effective one dimensional potentials given by $
V_{pp'}(y-y')={a^2}\sum_{n,n'}U(\Delta x_{nn'}, \Delta y_{pp'})|\phi(x_n)|^2|\phi(x_{n'})|^2$; 
and $\Delta x_{n,n'}=a(n-n'+{1\over2}(2y/3d-2y'/3d)_{\rm mod 2})$,
$\Delta y_{p,p'}=y-y'+pd\delta_{p,-p'}$ and $p=+$ ($p=-$) 
labels the sublattice $A$ ($B$). An analysis of the scattering processes shows that at low energies, small momentum transfer processes (long-range part of Coulomb interaction) dominate for reasonable ribbon widths ($M \geq 5$), similarly to carbon nanotubes \cite{fisher,gogolin}. These forward scattering terms include inter-($\Delta y_{p,-p}=y-y'+pd$) and intra-sublattices processes ($\Delta y_{p,p}=y-y'$). In terms of $\rho_p(y)=\sum_s\psi^{\dag}_{p,s}\psi_{p,s}$, (electron density at sublattice $p$), the Hamiltonian reads: 
\be
H=\int_0^L dy [ \frac{V}{2}\rho(y)\rho(y) - {\delta V\over2}\sum_{p}\rho_{p}(y)\rho_{-p}(y)].
\label{Hs}
\ee
where $\rho=\rho_A+\rho_B$, $V={a\sqrt{3}\over2}\sum_mV_{p,p}(y_m)$ and 
$\delta V={a\sqrt{3}\over2}\sum_m(V_{p,-p}(y_m)-V_{p,p}(y_m))$ measures the difference between intra- and inter-sublattice interactions.

\begin{figure}
\psfrag{W/a}{{ $W/a$}}
\psfrag{kc}{{ $K_c$}}
\psfrag{ks}{{ $K_{\sigma}$}}
\psfrag{d}{{ $\Delta~(eV)$}}
\includegraphics[width=.5\textwidth]{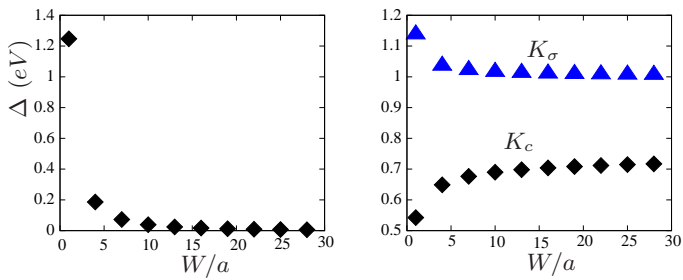}
\caption{Left: Charge band-gap of an armchair ribbon of length $L\simeq 400 \mu m$ as function
    of its width  $W$. The data show behavior $\Delta \approx 1/W^{\alpha}$ with  $\alpha = 1.6 $ (see discussion in text). 
   Right: $K_c$ and $K_{\s}$ as function of the ribbon's width.}
\label{kvscs}
\end{figure}

To solve the above Hamiltonian we use bosonization \cite{bbook}. As usual, bosonic fields are introduced $:
\psi_{r,s}={\eta_{r,s}\over\sqrt{2\pi a_0}}\exp[-ik_Fry-i{\sqrt{\pi/2}}(\Phi_c+s\Phi_{\s}+r\Theta_c+rs\Theta_{\s})   ]$
where $\eta$ is a Klein  factor, $\Phi_c$ and $\Phi_{\s}$ are 
charge and spin bosonic fields ($\Theta_c$ and $\Theta_{\s}$ are their 
duals). In terms of the total charge density ($\rho=\frac{\p_y\Phi_c}{\sqrt{2\pi}}$) and the spin current ($J=\frac{\p_y\Theta_{\s}}{\sqrt{2\pi}}$) the Hamiltonian $H(\Phi, \Theta)$ reads:
\bea
H(\Phi, \Theta)=\frac{v_{c/\s}}{2}
\int dy\sum_{c,\s}\big[\frac{(\p_y\Phi_{c/{\s}})^2}{K_{c/\s}}
+K_{c/\s}(\p_y\Theta_{c/{\s}})^2\big]&& \nn
-{\delta V\over 4\pi^2a^2}\int dy[\cos(\sqrt{8\pi}\Phi_c+4k_Fy)-\cos\sqrt{8\pi}\Phi_{\s}]~~&&
\label{Hboson1}
\eea
where $v_{c/\s}=v/K_{c,\s}$, $K_c=1/\sqrt{1+{V-3\delta V\over2\pi v}}$ and
$K_{\s}=1/\sqrt{1-{\delta V\over2\pi v}}$. This Hamiltonian describes
 two decoupled sine-Gordon models for charge and spin sectors. Notice that within this approximation,  metallic ribbons of width $W=a$ reduce to quantum spin chains \cite{bbook}. Analysis of Eq.\ (\ref{Hboson1}) shows that away from half-filling ($k_Fa=\pi/2$), spin and charge sectors are gapless. At half-filling, $K_{\s}>1$ and the spin field $\Phi_{\s}$ remains gapless, but $K_c <1$ and the charge field becomes gapped. The gap $\Delta$ is estimated in the self-consistent harmonic approximation \cite{bbook}:
\be
\Delta={v_c\over a} \left({\delta V\over 2\pi v_c} \right)^{1/2(1-K_c)}
\ee
Figure \ref{kvscs} shows the dependence of the band-gap and the $K_{c}, K_{\s}$ parameters on the ribbon width for a ribbon length $L=10^{6}\sqrt{3}a\simeq 400 \mu m$ (the dependence on the ribbon length enters in the definitions of $V$ and $\delta V$ as shown above). The band-gap takes sizable values for narrow ribbons 
($\Delta \simeq 130 meV$ for a $W \simeq 5 nm$ ribbon), although it decreases dramatically with ribbon width as $\Delta \approx 1/W^{\alpha}$, with $\alpha \to 1$ in the infinite length limit. The logarithmic dependence of $\alpha$ on $L$ results from the approximation of keeping only long-range scattering terms and can be dealt with the formalism presented in a systematic way. These results are in qualitative agreement with numerical calculations \cite{son} and experimental measurements \cite{kim2, avouris}. Recent DFT calculations \cite{son} on armchair ribbons  have found a width dependent gap that vanishes as $1/W$ in the limit of wide ribbons. The various widths considered in that case included insulating and
metallic ribbons, thus comparison with our results applies only for metallic ribbons  (see Ref.\cite{son}, Fig. 2b, N=3p+2). Interestingly, although DFT calculations include effects of electron-electron interactions, the opening of a gap in Ref. \cite{son} was associated with relaxed atomic positions at the edge with carbon atoms passivated by H atoms. Measurements on 
lithographically patterned graphene ribbons with different widths have also revealed non-metallic behavior with a width-dependent gap, explained as a consequence of quantum confinement. Comparisons of these results with theoretical predictions however remain at a qualitative level due to the undetermined nature of ribbon's edges in the samples used.

Unlike numerical methods, the approach used provides expressions for various correlation functions $\langle T_{\tau}{\cal O}(\tau,y){\cal O}^{\dag}(0,0)\rangle$, where ${\cal O}$ is an operator representing relative inter-sublattice charge (${\cal O}_{C}=\rho_A-\rho_B$) or spin (${\cal O}_{S^i}=S^i_A-S^i_B, i=x,y,z$) densities. 
These have the following expressions:
\bea
&\langle {\cal O}_{C}{\cal O}_{C} \rangle=N\langle \sin(y' + \sqrt{2\pi}\Phi_c)\cos(\sqrt{2\pi}\Phi_{\s}) \rangle&\nn
&\langle {\cal O}_{S^z}{\cal O}_{S^z} \rangle=N\langle \cos(y' + \sqrt{2\pi}\Phi_c)\sin(\sqrt{2\pi}\Phi_{\s}) \rangle&\nn
&\langle {\cal O}_{S^{\pm}}{\cal O}_{S^{\pm}} \rangle=N\langle \cos(y' + \sqrt{2\pi}\Phi_c)e^{\pm i(\sqrt{2\pi}\Theta_{\s})}\rangle &
\label{corrfunc}
\eea
with $y' = 2k_Fy$ and $N=\frac{2}{\pi a}$. At half-filling, the charge field is locked at $\Phi_c=0$ and the dominant correlation functions are $\langle {\cal O}_{S^z}{\cal O}_{S^z} \rangle \sim y^{-K_{\s}}$ and $\langle {\cal O}_{S^i}{\cal O}_{S^i} \rangle \sim y^{-1/K_{\s}}$ ($i=x,y$). 
These results suggest enhanced magnetic correlations in the direction parallel to the graphene plane for narrow ribbons. Away from half-filling however, the dominant correlations depend on the values of $K_{c}$ and $K_{\sigma}$ which are length dependent in the formalism presented. For ${K_{\s}\over3} \le K_c \le 1$ the most relevant operators are ${\cal O}_C,{\cal O}_{S^{x,y}}$ with correlation functions decaying algebraically as $\langle {\cal O}_{C}{\cal O}_{C} \rangle \sim y^{-K_c-K_{\s}}$ and $\langle {\cal O}_{S^{x,y}}{\cal O}_{S^{x,y}} \rangle \sim y^{-K_c-1/K_{\s}}$; respectively. In the regime $K_c \le {K_{\s}\over 3}$ the dominant correlation function is  ${\cal O}_{CC}=(\rho_A-\rho_B)(\rho_A-\rho_B)$ that decays as $\langle {\cal O}_{CC}{\cal O}_{CC} \rangle \sim y^{-4K_c}$. This regime might be achieved only for narrow ribbons. 

\begin{figure}
\psfrag{Probability}{Probability}
\psfrag{0}{0}\psfrag{1}{1}\psfrag{2}{2}\psfrag{3}{3}\psfrag{4}{4}
\psfrag{W/a}{ $W/a$}
\psfrag{p1}{ $\re^{i\theta_0}$}
\psfrag{p2}{ $\re^{i\theta_+}$}
\psfrag{p3}{ $\re^{i\theta_-}$}
\includegraphics[width=.5\textwidth]{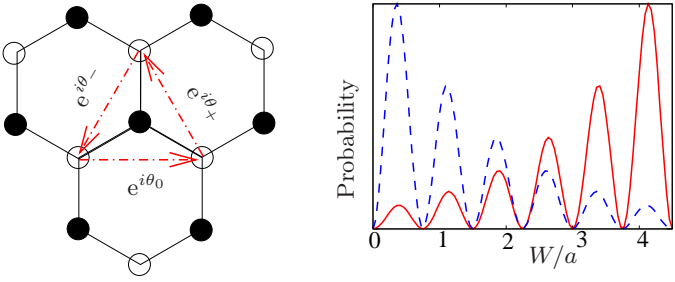}
\caption{Left: I-SO term with second-neighbor hopping of an electron in sub-lattice $A$, coupled to the spin of an electron on $B$. Eq.(\ref{soterm}) results from adding clockwise motion and exchanging the role of  both sublattices. Here  $\theta_0=k_xa,~\theta_\pm=-{k_xa/2}\pm{k_ya/2\sqrt{3}}$. Right: probability amplitude for a right mover with spin-down (dash-line) and with spin-up (solid-line) at the edges of the ribbon as a function of ribbon width.}
\label{spcu}
\end{figure}

{\it Effect of intrinsic spin-orbit (I-SO) interaction.}
Another interesting characteristic of graphene is that its lattice symmetry supports a particular spin-dependent second-neighbor hopping interaction, known as the I-SO interaction. This interaction is responsible for the appearence of spin-filtered edge states \cite{kanemele, macdonald} which have received much attention because of their role in spin quantum Hall physics and their relevance for spintronic circuitry.
As proposed in \cite{kanemele}, the I-SO interaction is an extension of Haldane's original model \cite{haldane} for the quantum Hall effect with total zero magnetic flux that includes both spins. Its Hamiltonian representation is given by: 
\begin{equation}
H_{SO}  =  \left( \begin{array}{cc}
                              \gamma \sigma^z & 0  \\
			      0      & -\gamma \sigma^z  \end{array}\right)
\label{soterm}
\end{equation}			      
where $\gamma =   2 t'(\sin(q_xa)-2\sin(q_xa/2)\cos(\sqrt{3}k_ya/2))$, $t'$ is the I-SO coupling, and $\sigma^z$ is the spin Pauli matrix. An intuitive picture for this term is shown in Fig.\ \ref{spcu}: the orbital motion of an electron hopping in sublattice $A (B)$ couples to the spin of an electron sitting in the enclosed site of sublattice $B (A)$. Since the hopping is spin-dependent, the interaction preserves time reversal invariance. After expanding the Hamiltonian around the $(K, K')$ points, a linear dispersion mode along the $y$-direction is found: 
\bea
\Psi_{\mp}^s = N \sin{4\pi x\over 3a}
\re^{\mp s\gamma_0 (x-W'/2)}{\re^{ik_yy}\over\sqrt{2L}}\left( \begin{array}{c}
                              1\\
			      \mp i    \end{array}\right) \; ,
\label{freewavefunction}
\eea
 where $N= \sqrt{2\gamma_0\over \sinh\gamma_0 W'}$ is the normalization factor, $\gamma_0=3\sqrt{3}t'$, and $s=\pm$ labels the spin. This expression reduces to the known result in the limit of semi-infinite ribbons \cite{sengupta} and it is in sharp contrast with predicted results on graphene sheets \cite{yao}. As Eq.\ (\ref{freewavefunction}) indicates, these new modes are localized on the edges and are spin-filtered states: the probability currents along the edges are spin-polarized \cite{kanemele}. The right panel in Fig.\ \ref{spcu} shows an example of such state. The velocity of this linear mode is equal to the Fermi velocity of the free system since in the presence of the I-SO term, $k_{x}$ acquires an imaginary part $ k_{x} = 4\pi/(3a) + i \gamma_{0}$ that preserves the linear dispersion relation. 
As done previously, it is possible to define an effective Hamiltonian along the $y$-direction and write it in terms of new boson fields $(\Phi_c', \Phi_{\sigma}')$ and their duals.  
By including only forward scattering terms, the effect of Coulomb interactions on these states is obtained from :
\be
H = H(\Phi', \Theta') + \int dy \frac{V^{1,2}-2\delta V^{1,2}}{2\pi} \p_y \Phi_{c}' \p_y\Theta_{\s}' 
\label{hamiso}
\ee
with the new velocities and Luttinger parameters:
\bea
&&v_{c}'=v/K_{c}'\;\;\;K_c'=1/\sqrt{1+({V^{1,1}-2\delta V^{1,1}-\delta V)/2\pi v}}\nn
&&v_{\s}'=v\sqrt{(1+{V^{2,2}- 2\delta V^{2,2}\over 2\pi v})
(1-{\delta V\over 2\pi v})}\nn
&&K_{\s}'=\sqrt{(1+{V^{2,2}- 2\delta V^{2,2}\over 2\pi v})/(1-{\delta V\over 2\pi v})}  \; .
\eea
Here we have defined
$V^{j,k}={a\sqrt{3}\over2}\sum_mV_{p,p}^{j,k}(y_m)$ and
$\delta V^{j,k}={a\sqrt{3}\over2}\sum_m(V_{p,-p}^{j,k}(y_m)-V_{p,p}^{j,k}(y_m))$ where
\bea
V_{pp'}^{jk}(y-y')={a^2}\sum_{n,n'}U(\Delta x_{nn'}, \Delta y_{pp'})\cosh(\alpha_{n}^{j})&&\nn
\cosh(\alpha_{n'}^{k})(-i)^{j+k}| {2\gamma_{0}\over \sinh\gamma_{0} W'}
\sin{4\pi x_n\over 3a}\sin{4\pi x_{n'}\over 3a}&&
\eea
with $\alpha_{n}^{l}=2\gamma_{0} x_n-\gamma_{0} W'+i{l\pi\over2}$.
As Eq.\ (\ref{hamiso}) shows, Coulomb interactions renormalize the bare couplings and introduce a new term $ \p_y\Phi_c'\p_y\Theta_{\s}'$ that destroys spin-charge separation. This term respects time reversal invariance and is present in quantum wires with Rashba spin-orbit interactions \cite{mor, iu}. A rotation by an angle $\eta$ diagonalizes the quadratic terms of the Hamiltonian:
\begin{equation}
\tan2\eta={\lambda\over v_{c}^2-v_{\s}^2};\;\;\lambda=v(1-{\delta V\over 2\pi v}){V^{1,2}-2\delta V^{1,2}\over\pi}
\end{equation}
This rotation modifies the last term in Eq.\ (\ref{hamiso}) producing corresponding changes in correlation functions. As it has been proposed \cite{yao, str}, the strength of the I-SO interaction is small compared to the bandwidth ($t \simeq 3eV, t' \simeq 0.001 meV$), implying that the rotation angle $\eta \ll 1$. As a consequence, the strong coupling physics of Eq.\ (\ref{hamiso}) and the conclusions of the previous section remain unchanged in the presence of I-SO.

To summarize, we studied the low-energy physics of armchair nanoribbons in the presence of intrinsic spin-orbit and electron-electron interactions within a Dirac model. We focused on states with linear dispersion near the neutrality (Dirac) points, and showed that they persist in the presence of I-SO interactions, as spin-filtered states localized on the ribbon's edges. These results have direct testable experimental consequences and are in contrast with predictions of insulating behavior caused by this particular interaction in graphene planes. For half-filled systems, we showed that small-momentum transfer processes, due to the long-range part of Coulomb interactions, open a charge-gap in the spectrum, while keeping the spin sector gapless. The gap $\Delta$ is strongly dependent on the ribbon's width and vanishes in the limit of infinite ribbons as $1/W$ in qualitative agreement with DFT calculations. Unlike numerical methods \cite{son}, the formalism used provides clarification on the origin of the gap and gives an explicit expression for its dependence on the ribbon width as well as expressions for various correlations functions. Further analysis of the results suggests an enhanced magnetic correlation for narrow ribbons in qualitative agreement with interpretations proposed for recent experimental measurements \cite{abanin}. For graphene ribbons away from half-filling, charge and spin sectors remain gapless even in the presence of interactions.
These results also hold for weak I-SO interactions.

We acknowledge discussions with C. Busser, M.A. Barral, S.E. Ulloa and H. Fertig. This work was partially supported by NSF-NIRT No 0304314.

\end{document}